\documentclass[runningheads]{svmult}
\usepackage{makeidx}   
\usepackage{graphicx}  
\usepackage{subeqnar}  
\usepackage{multicol}  
\usepackage{cropmark} 
\usepackage{physprbb}  
\begin{document}
\title*{The CRESST Dark Matter Search}
\toctitle{The CRESST Dark Matter Search}
%
%
\titlerunning{The CRESST Dark Matter Search}
%
\author{J.~Jochum\inst{1}
\and M.~Bravin\inst{2}
\and M.~Bruckmayer\inst{2}
\and C.~Bucci\inst{3}
\and S.~Cooper\inst{4}
\and C.~Cozzini\inst{2}
\and P.~DiStefano\inst{2}
\and F.~v.Feilitzsch\inst{1}
\and T.~Frank\inst{2}
\and T.~Jagemann\inst{1}
\and R.~Keeling\inst{4}
\and H.~Kraus\inst{4}
\and J.~Lush\inst{4}
\and J.~Marchese\inst{4}
\and O.~Meier\inst{2}
\and P.~Meunier\inst{2}
\and U.~Nagel\inst{1}
\and D.~Pergolesi\inst{2}
\and F.~Pr\"obst\inst{2}
\and Y.~Ramachers\inst{4}
\and J.~Schnagl\inst{1}
\and W.~Seidel\inst{2}
\and I.~Sergeyev\inst{2}
\and M.~Stark\inst{1}
\and L.~Stodolsky\inst{2}
\and S.~Uchaikin\inst{2}
\and H.~Wulandari\inst{1}}
\authorrunning{J.~Jochum CRESST collabroation}
%
%
\institute{Technical University Munich,  
		James Franck Str.1, 85747 Garching, Germany
\and Max Planck Institute for Physics, F\"ohringer Ring 6, 80805 Munich, Germany
\and Laboratori Nationali Del Gran Sasso, Strada Str. 17/bis, 67010 Assergi, Italy
\and Oxford University, Keble Road, Oxford OX1 3RH, UK}

\maketitle              

\begin{abstract}
We 
present
the current status of CRESST
(Cryogenic Rare Event Search using Superconducting
Thermometers)
project and new results concerning
detector development.  The basic
technique involved is to search for WIMPs by the
measurement  of non-thermal
phonons, as created by WIMP-induced nuclear recoils.  
Combined with our newly developed method for the 
simultaneous measurement of
scintillation light,  strong  background discrimination is
possible, resulting in a substantial increase in WIMP detection
sensitivity. 
\end{abstract}

\section{CRESST and the Dark Matter Problem}

After a long period of development, cryogenic detectors are now
coming on line and deliver
significant results in particle-astrophysics and weak interactions. 
The goal of the  CRESST project  is the direct detection
of elementary particle dark matter and the elucidation of its
nature. Particle physics provides a well motivated dark matter candidate  through the lightest supersymmetric (SUSY) particle, the
`neutralino' and one can find candidates in a wide mass range \cite{pok}.
Generically, such particles are called WIMPS (Weakly Interacting
Massive Particles). WIMPS are expected to
interact with ordinary matter by elastic scattering on nuclei. 
Conventional methods for  direct detection  rely on the ionization
or scintillation caused by the recoiling nucleus. This leads to limitations
connected with the relatively high energy involved in producing electron-hole pairs. Cryogenic detectors use the much lower energy
excitations, such as phonons.
Since the principal physical effect of a WIMP nuclear recoil is the
generation of phonons, cryogenic calorimeters are well suited for WIMP detection.  Further, when this technology  is
combined with charge or light detection the resulting background suppression leads to a powerful technique to search for the rare nuclear recoils. 

The detectors developed by the CRESST collaboration consist of a
dielectric target-crystal with a small  superconducting film evaporated onto the surface. When this film is held at a temperature
in the middle of its superconducting  to normal conducting phase transition, it   
functions as a highly sensitive thermometer. The detectors presently employed in
Gran Sasso use tungsten (W) films and sapphire ($Al_2O_3$)
absorbers, running near 15 mK. The technique can also be applied to a variety of
other materials.
The small change in  temperature of the superconducting  film resulting from
an energy deposit in the target leads to a relatively large
change in the film's resistance. This change in resistance 
is measured with a SQUID. 
A small separate detector of the same type is used to see the light emitted  when the target is a scintillating crystal.

\section{Present Status of CRESST}

The task set for the first stage of CRESST was to show the operation of four 262 g sapphire detectors, with a threshold of 500 eV under low background conditions. Meeting this goal involved  the setting up of a low background, large volume, cryogenic installation and the development of massive, low background detectors with low energy thresholds.

The central part of the CRESST installation at the LNGS is the cryostat. The cryostat design separates the detectors in the "cold box" from the dilution refrigerator by a 1.5 m long "cold finger", with internal cold lead shielding blocking the line-of-sight to the detectors. The cold box is  constructed entirely of low background materials, without any compromise. It is surrounded by shielding consisting of 20 cm of lead and 14 cm of copper. The volume for the detectors of about 30 l is large enough to use much larger detectors in the future. The cold box and  shielding are installed in a clean room area with a measured clean room class of 100. For servicing, the top of the cryostat can be accessed from the first floor outside the clean room.  

The installation was completed at Gran Sasso in 1997 and a series of detector tests were made in the prototype cold box during 1998. The purpose of the prototype cold box was to test the mechanical and cryogenic functioning of the design and to provide a reasonably shielded environment for completing the development of the 262 g sapphire detectors. As can be expected in such a complicated setup, the first runs showed that several details needed improvement: Vibrations caused by the needle valve of the dilution refrigerator's 1 K pot needed to be eliminated by installing a fixed impedance; the design of the detector holders needed to be improved to reduce the vibrational effect of the boiling of the liquid nitrogen; the slow heat release due to enclosed H2  in the Cu of some new components needed to be understood and eliminated by carefully selecting  low background Cu which does not have this problem. 

To allow monitoring the longterm stability in a search for dark matter we have developed W thermometers with attached electrical heaters. A periodical injection of heater pulses of a number of different energies allows to precisely monitor the stability of the energy calibration of the detectors and also to correct for possible deviations from linearity. 

\begin{figure}[htb]
\begin{center}
\rotatebox{90}{\includegraphics[width=.6\textwidth]{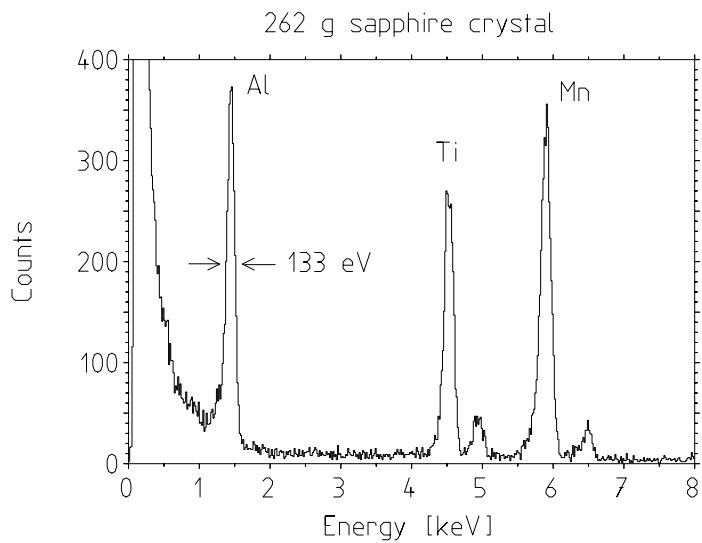}}
\end{center}
\caption{Low energy X-ray spectrum from one of the 262gr Sapphire-detectors. The energy resolution at 1keV is 133eV FWHM. The energy threshold is below 500eV. The rise in countrate at low energies is due to Auger electrons.}
\label{fig:133eV}
\end{figure}

At the end of this testing periods we had four 262 sapphire detectors which achieved energy resolutions in the range of 200 eV (FWHM) at 1.5 keV. The best detector reached a resolution of 133 eV at 1.5 keV. The spectrum is shown in fig.~\ref{fig:133eV}. As mentioned above, these tests were made in the prototype cold box, which was made of the right type of copper but which had been exposed at the surface for many years with resultantly large contamination due to activation by cosmic rays. Also no attempt was made after machining to remove surface impurities. In order to replace it at the first opportunity, detector tests were stopped in spring 1998 and efforts concentrated on preparing the new cold box. 

The new low background cold box was made using copper which had been stored underground since production and between machining steps. After machining the surface of each piece was cleaned by electropolishing and subsequent rinsing with high purity water. The pieces were then stored in a gastight transport container made of PE and flushed with nitrogen. 

Our clean room in Gran Sasso was improved so that during the whole installation of the new cold box in 1998 a level better than class 100 was reached. During 1999, a  series of  first measurements with four  262g detectors under low background conditions was performed in the new cold box.  The measured  rate was of the order of a few 10 counts/ (kg keV day)  above 30 keV and below 1 count/(kg keV day) above 100 keV. This was much larger than expected and not caused by radioactivity. The detectors are mounted facing each other with no material in between. The complete absence of any events in true time coincidence between any pair of the detectors excludes surface contamination of the crystal with $\beta$-emitters and contamination with $\gamma$-emitters in the crystals and the surrounding. Also $\beta$-emitters within the crystal were most likely excluded by the shape of the measured energy spectra. Moreover, the rate had a nonpoissonian character in time, which can not be caused by radioactivity. 

 In subsequent runs the detectors have been insulated from external vibrations by mounting them on a spring-suspended platform (horizontal and vertical resonance frequencies of 1.5 Hz and 3 Hz). These steps lead to an   immunity of the base line signal against vibrations in tests where external vibrations were applied to the cryostat. Nevertheless the rate did not decrease. This lead us to exclude vibrations as the origin of the background. The idea of using a spring suspended  platform was motivated by the very good results of the Milano group with a similar type of mounting. 

In  a further run we were investigating the possibility that electromagnetic interference with a too short duration to be seen directly by the SQUID might heat the W-thermometer and cause thermal signals with a shape very similar to particle pulses. A modification of the readout circuit of one detector in order to suppress the heating effect of  interference pulses made the system completely immune to any external electromagnetic interference created for test purposes inside the faraday cage. However, the background rate did not decrease and we exclude electromagnetic interference as the source of our background. 

As a further diagnostic tool a 262 g sapphire detector carrying two W thermometers was used. We found that the signals from both thermometers are strictly coincident in time with a constant ratio of the pulse heights and again with the same shape as particle pulses. Again the signals were not coincident with the signals of another detector. This clearly demonstrated that all signals originate from energy depositions which create high frequency phonons in the sapphire and confirms again that the background is not due to electrical interference. 

After having excluded radioactivity, vibrations and electronic interference as the source of the background, a remaining possibility was 
that the energy depositions in the crystals are due to random structural relaxation in the supporting structure of the crystal. 
This possibility was investigated in spring 2000. The sapphire spheres, supporting the absorber crystal, were replaced by small Teflon mounts. 
This way the contact surface of the crystal to its holders was enlarged and the pressure reduced. 
This measure resulted in a significant reduction of the background rate. 
For energies from 15 keV to 25 keV the rate is now below 1 count / (kg keV day), which is in the range we are aiming for. At lower energies the rate is dominated by a line at 8keV, presumably the K$_{\alpha}$ X-ray line of Copper. 
We are confident of being able to present results on Dark Matter during the year 2000.

\section{The new Detectors}

If all disturbances are removed, the remaining background  will be dominated by
$\beta$ and $\gamma$ emissions from nearby
radioactive contaminants. These produce exclusively  electron
recoils in the detector. 
Therefore, dramatic improvements in sensitivity  are to be expected
if the detector itself is capable of distinguishing  electrons from
nuclear recoils and rejecting them. 
We have recently developed a system, presently using CaWO$_4$
crystals as the absorber,  where a measurement of scintillation 
light is carried out in parallel to the phonon
detection. We find that these devices clearly discriminate 
nuclear recoils from electron recoils. 

\begin{figure}[htb]
\begin{center}
\hskip-2.5truecm{\includegraphics[width=.6\textwidth]{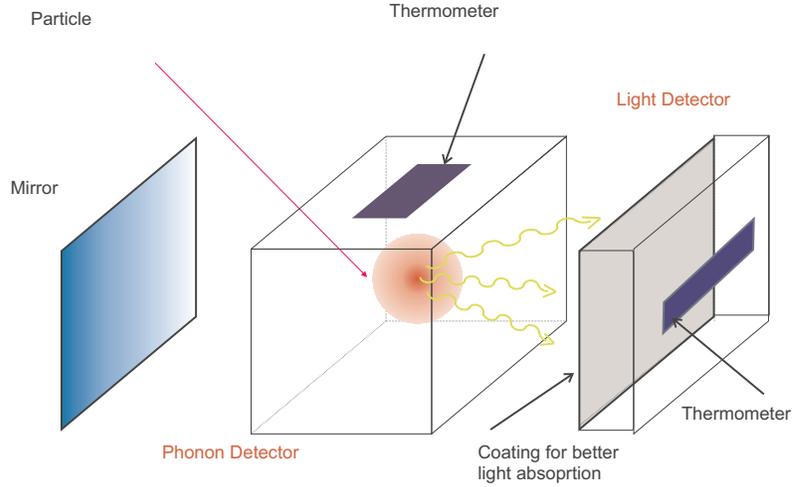}}
\end{center}
\caption{Schematic view of the arrangement used for the simultaneous light and 
phonon detection. A second smaller calorimeter next to the absorber is used to measure the scintillation light. Mirrors or a reflective coating on the inside of the detector holders improve the light collection efficiency. The light detector is coated for 
better light absorption.}
\label{fig:schema}
\end{figure}

\begin{figure}[tb]
\begin{center}
\includegraphics[width=.6\textwidth]{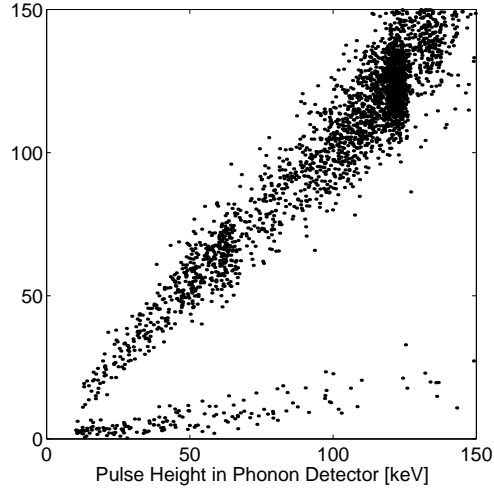}
\end{center}

\caption{Pulse height in the light detector versus pulse height in
the phonon detector. The  scatter plot has been  measured with 
an electron-, a photon-, and a neutron source. The upper line corresponds to electron recoils, the lower to nuclear recoils. On can clearly discriminate the two down to energies of about 10keV.}
\label{fig:lpwithn}
\end{figure}

 We first studied the light output of several intrinsic scintillating crystals. Al crystals tested so far (BGO, BaF$_2$, PbWO$_4$, CaWO$_4$) showed adequate light output at mK temperatures. The high sensitivity of the W phase transition thermometers allows us to use a small cryogenic calorimeter to measure the light  emitted  from a scintillating absorber crystal. To demonstrate the detection principle we developed a test detector consisting of a scintillating CaWO$_4$ absorber with a mass of 6 g and a separate light detector.  Both the absorber and the light detector were instrumented with their own tungsten superconducting phase transition thermometer operating at about 12mK. 
The system is shown schematically in fig.~\ref{fig:schema} .  
It consists of two independent detectors:
A scintillating  absorber with a superconducting phase
transition thermometer on it, and a similar but smaller detector placed next 
to it to detect the scintillation light from the first detector. 
A detailed description is given in \cite{cawopaper}. 
With this device we succeeded to simultaneously measure the phonons and the scintillation light from particle interactions in the CaWO$_4$ absorber. The detector was irradiated with  
$\gamma$-rays and electrons. Adding an external neutron source we could demonstrate clear discrimination between electron and nuclear recoils. 

Figure ~\ref{fig:lpwithn} shows a scatter plot of the pulse 
heights observed in the light detector versus
the pulse height observed in the phonon detector. A clear 
correlation between the light and phonon signals is observed. 
A second line can be seen  due to nuclear recoils induced by neutrons from an Americium-Beryllium source. It
is to be observed that electron
and nuclear recoils can be clearly distinguished down to a
threshold of 10keV. A detailed evaluation yields a background rejection factor of 98\% in the energy range between 10\,keV and 20\,keV, 99.7\% in the range between 15\,keV and 25\,keV and better than 99.9\% above 20\,keV. 

In Munich we are presently working on the development of a 300 g CaWO$_4$ prototype detector. We recently demonstrated a good light collection with such a 300 g crystal, sufficient to realize a prototype detector for Gran Sasso. We plan to install such a 300 g detector this year in Gran Sasso. Projections indicate that this step would move CRESST from its good  sensitivity to low mass dark matter WIMPs  with the sapphire detectors to a position where it can compete favorably in the higher mass regime with the best proposed experiments.

\section{Next Steps for CRESST}

\begin{figure}[t]
\begin{center}
\rotatebox{90}{\includegraphics[width=.64\textwidth]{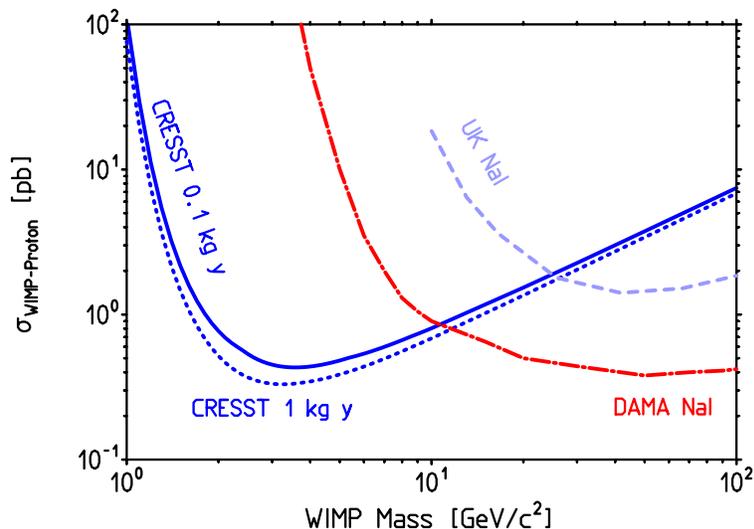}}
\end{center}
\caption[]{
Equivalent WIMP-proton cross section limits (90\% CL) for spin
dependent
interactions as a function of the WIMP-mass, as expected for the
present
CRESST sapphire detectors with a total mass of 1 kg. The
expectation is based
on a threshold of 0.5 keV, a background of 1\,count/(kg\,keV\,day) and
an exposure
of 0.1 and 1\,kg year. For comparison the present limits from the
DAMA
\cite{rita96} and UKDMC \cite{ukdm} NaI experiments are also
shown.}
\label{fig:spindep}
\end{figure}

Due to the complementary detector concepts of low threshold
calorimeters on the one hand and detectors with the simultaneous
measurement of light and phonons on the other, CRESST can cover a
very wide range of WIMP masses.
The present sapphire detectors, with their extremely low energy
thresholds and a low mass target nucleus with high spin ($Al$),  cover the
low WIMP mass range from 1 GeV to 10 GeV in the sense that
they are presently the only detector type able to explore this mass range effectively. Besides CRESST, also the ROSEBUD-Collaboration is 
running very similar cryogenic sapphire detectors \cite{cebrian}. 
Data-taking with the present sapphire ($Al_2O_3$) detectors
(262\,g each) will continue during 2000. Possible limits for spin dependent interaction are shown in fig. ~\ref{fig:spindep}.

\begin{figure}[t]
\begin{center}
\rotatebox{90}{\includegraphics[width=.70\textwidth]{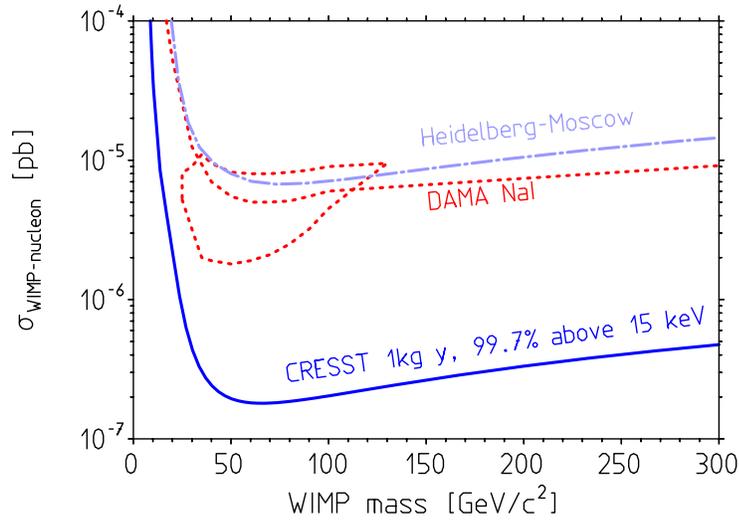}}
\end{center}
\caption[]{WIMP-nucleon cross section limits (90\% CL) for scalar (coherent)
interactions
as a function of the WIMP mass, expected for a 1\,kg CaWO$_4$ detector
with a background rejection of 99.7\% above a threshold of 15 keV
detector and  1 year
of measurement time in the CRESST set-up in Gran Sasso.
For comparison the limit from the
Heidelberg-Moscow $^{76}$Ge
experiment \cite{hdmnew} and
the DAMA NaI limits \cite{rita96} (with the contour for positive
evidence
\cite{ritapositive}) is also shown.}
\label{fig:kurzfr}
\end{figure}

The detectors with  background suppression using the simultaneous measurement of
scintillation light and phonons will have target
nuclei of large atomic number, such as tungsten,  making  them particularly sensitive to  WIMPs with coherent interactions and higher mass above about 20GeV. 
In 2000 we intend the first installation of this next detector generation at Gran Sasso. 
A 60 GeV WIMP with the 
cross section claimed in \cite{ritapositive}
would give about 55 counts between
15 and 25 keV in 1\,kg CaWO$_4$ within one year. A background of
1~count/(kg\,keV\,day) suppressed with 99.7\% 
would leave 11 background counts in the same energy range.
A 1\,kg CaWO$_4$ detector with 1 year of measuring time in the
present setup of CRESST 
should thus allow a comfortable test of  the reported positive
signal as shown in fig ~\ref{fig:kurzfr}.

\section{Long Term Perspectives}

After successful implementation of the first CaWO$_4$ detectors we intend to upgrade the multichannel SQUID read out and
systematically increase the detector mass, which can go up to about
100\,kg . 

\begin{figure}[htb]
\begin{center}
\includegraphics[width=.75\textwidth]{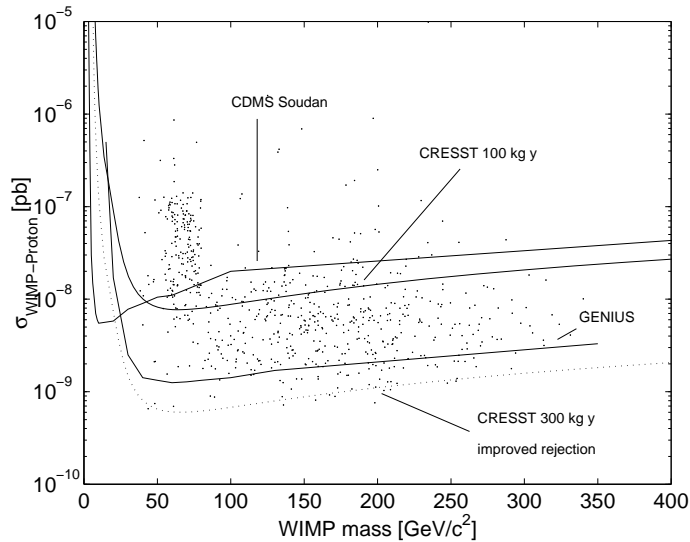}
\end{center}
\caption[]{WIMP-nucleon cross section limits (90\% CL) for coherent 
interactions,  as a function of the WIMP mass, expected for a CaWO$_4$ detector
with a background of 1~count/(keV kg day) , a background suppression of
99.9\% above
a threshold of 15 keV,  and an exposure of 100\,kg-years in the
CRESST setup.
With a suppression of 99.99\% above 15 keV, a reduced background of
0.1 counts/(kg keV day), and an increased exposure of 300 kg
years
most of the MSSM parameter space would be covered.
For comparison, the projected sensitivity of CDMS at Soudan
\cite{nam}, and of the GENIUS experiment
\cite{Genius} are also shown. All sensitivities are 
scaled to a galactic WIMP density of 0.3 GeV/cm$^3$. 
The dots represent expectations for WIMP-neutralinos
calculated
in the MSSM framework with non-universal scalar mass unification
\cite{bednyakov}.}
\label{fig:langfr}
\end{figure}

With a 100 kg CaW0$_4$ detector and a background level of 1count/kg/kev/day,  
the sensitivity  shown in fig.~\ref{fig:langfr} can be reached in
one year of measuring time.
If we wish to  cover most of the MSSM parameter space of SUSY with
neutralino dark matter,
the exposure would have to be increased to about 300 kg years, the
background suppression improved to
about 99.99 \% above 15 keV, and the background lowered to 0.1 count/(kg keV
day).  The recent tests in Munich with  CaWO$_4$, which were limited by ambient 
neutrons, 
suggest that a suppression factor of this order should be within reach
underground,  with the neutrons well shielded and employing a muon veto.  

If WIMPs are not found, at some point  the  neutron flux, which
also gives nuclear recoils, will begin to limit
further improvement.
With careful shielding the  neutron flux in  Gran Sasso should
not limit the sensitivity within the exposures assumed for the
upper CRESST curve in fig.~\ref{fig:langfr}.
With still larger exposures, the  neutron background  may still be
discriminated against large mass WIMPS. This can be done by
comparing  different target materials, which is possible with the
CRESST technology, since different variations with nuclear number
for the  recoil spectra are to be expected with different mass
projectiles.

\section{Conclusions}

The installation of the large volume, low background, cryogenic facility of 
CRESST at the Gran Sasso Laboratory is completed. 
The highly sensitive CRESST sapphire will allow to explore the low mass WIMP range. 
The new detectors with the simultaneous measurement of phonons and 
scintillation light allow to distinguish the nuclear recoils very effectively  
from the electron recoils caused by background radioactivity. 
For medium and high mass WIMPs this results in one of the highest sensitivities 
possible with today's technology.  
The excellent background suppression of cryodetectors with active
background rejection makes them much less
susceptible to systematic uncertainties than conventional
detectors,
which must rely  heavily on a subtraction of radioactive
backgrounds.
Since this kind of systematic uncertainty cannot be compensated by
an
increase of detector mass, even moderate sized cryogenic detectors
can achieve much better sensitivity than large mass conventional
detectors.

\section{Acknowledgement }

This work was supported by the ``SFB 375-95 f\"ur Astro-Teilchenphysik der DFG and the EU research network contract number FMRX-CT98-0167 (DG12-MIHT).

\end{document}